\documentclass[pre,twocolumn,superscriptaddress]{revtex4}
\usepackage{amsmath}
\usepackage{graphicx}

\input{tcilatex}

\begin{document}

\title{The Rise and Fall of a Networked Society}
\author{Matteo Marsili}
\affiliation{The Abdus Salam International Centre for Theoretical Physics, Strada
Costiera 11, 34014 Trieste Italy}
\email{marsili@ictp.trieste.it}
\author{Fernando Vega-Redondo}
\affiliation{Departamento de Fundamentos del An\'{a}lisis Econ\'{o}mico and Instituto
Valenciano de Investigaciones Econ\'{o}micas, Universidad de Alicante, 03071
Alicante, Spain}
\email{vega@merlin.fae.ua.es}
\author{Franti\v{s}ek Slanina}
\affiliation{Institute of Physics, Academy of Sciences of the Czech Republic,
Na~Slovance~2, CZ-18221~Praha, Czech Republic }
\email{slanina@fzu.cz}
\date{\today}

\begin{abstract}
We propose a simple model of the evolution of a social network which
involves local search and volatility (random decay of links). The model
captures the crucial role the network plays for information diffusion. This
is responsible for a feedback loop which results in a first-order phase
transition between a very sparse network regime and a highly-connected
phase. Phase coexistence and hysteresis take place for intermediate value of
parameters. We derive a mean-field theory which correctly reproduces this
behavior, including the distribution of degree connectivity and the
non-trivial clustering properties.
\end{abstract}

\maketitle

Recent phenomenological studies on complex networks in social sciences have
uncovered ubiquitous nontrivial statistical properties, such as scale free
distribution of connectivity or small world phenomena 
\cite{alb_bar_01,dor_men_01,wa_stro_98}. These properties have striking
consequences on the processes which take place on such networks, such as
percolation \cite{dor_men_sam_01a}, diffusion 
\cite{far_der_bar_vic_01,goh_kah_kim_01b}, phase transitions \cite{ising} and
epidemic spreading \cite{vespignani}. The research on complex networks
raises questions of a new type as it addresses phenomena where the topology
of interactions is part of the dynamic process. This contrasts with
traditional statistical mechanics, where the topology of the interaction is
fixed \emph{a priori} by the topology of the embedding space.

Phenomena of this type are quite common in social sciences where
agents purposefully establish cooperative links \cite{Putnam}. Links
between individuals in a social network support not only the
socioeconomic interactions that determine their payoffs, but it also
carries information about the state of the network. This aspect has
important consequences in the long run if the underlying environment
is volatile. In this case, former choices tend to become obsolete and
individuals must swiftly search for new opportunities to offset
negative events. The role of the network for information diffusion is
particularly apparent, for example, pertaining to the way in which
individuals find new job opportunities. It has been consistently shown
by sociologists and economists alike \cite{G,T} that personal
acquaintances or neighborhood effects play a prominent role in job
search. This, in turn, leads to significant correlation in employment
across friends, relatives, or neighbors. The common thesis proposed to
explain this evidence is that, in the presence of environmental
volatility, the quantity and quality of one's social links --
sometimes referred to as her \emph{social capital} \cite{Coleman}-- is
a key basis for search and adaptability to change.

The study of complex networks has been mainly concerned with simple
phenomenological models reproducing the main \emph{\ stylized
facts}. In contrast, the socioeconomic literature has studied
game-theoretic models of network formation \cite{JWol} in a fixed
environment. Here, our objective is to integrate and enrich both
approaches, proposing a stylized model of a society that embodies the
following three features: (i) agent interaction, (ii) search cum
adjustment, and (iii) volatility (i.e. random link removal).
Individuals are involved in bilateral interaction, as reflected by the
prevailing network. Through occasional update, some of the existing
links have their value deteriorate and are therefore lost. In
contrast, the individuals also receive opportunities to search that,
when successful, allow the establishment of fresh new links. Over
time, this leads to an evolving social network that is always adapting
to changing conditions.  The model aims to capture the continuous
struggle of search against volatility, which is at the heart of
network's dynamics. In the long run, the so-called Red Queen Principle
\cite{VV} applies: \textquotedblleft ...it takes all the running you
can do, to keep in the same place.\textquotedblright\ Agents'
continuous search must be strong enough to offset volatility if a
dense and effective social network is to be preserved. On the other
hand, search can be effective only in a densely networked society. So
information diffusion and a dense network of interactions are two
elements of a feedback self-reinforcing loop. As a result, the system
displays a discontinuous phase transition and hysteresis, enjoying
some resistance to a moderate deterioration of the underlying
environmental conditions. Such a resilience can be interpreted as
consequence of the buffer effects and enhanced flexibility enjoyed by
a society that has accumulated high levels of social capital. 

These features are captured by a mean field theory which is in good
agreement with numerical simulation results. This theory highlights
the particular role that clustering plays in the dynamics of the
model. Indeed search is particularly effective when clustering is
low whereas it is suppressed in a high clustered society. 

The model introduced here is a simplification of a more elaborate
model proposed by one of us in \cite{VR} to understand how the network
dynamics impinges on strategic behavior. The model is also similar to
that proposed in Ref. \cite{DEB} to explain the emergence of the
small-world property \cite{wa_stro_98} in social networks. There are,
however, important differences between \cite{wa_stro_98} and our
approach, as we shall discuss later at the end of the paper. Related
issues, within the vast recent literature on network dynamics, have
also been addressed in Ref. \cite{ebe_bor_02}, that studied the
evolution of network among agents involved in an iterated Prisoner's
Dilemma, and in Ref. \cite{der_far_pal_vic_03}, that found a
topological phase transition in networks that minimize a suitably
chosen cost function.

Formally, the network is given by a set of nodes $N$ and the
corresponding adjacency matrix $A$ with elements $A_{ij}\in \{0,1\}$,
for $i,j\in \{1,2,...,N\}$. The value $A_{ij}=1$ means that there is a
link connecting nodes $i$ and $j$, while $A_{ij}=0$ holds
otherwise. We will require $A_{ii}=0$ (no on-site loops) and
$A_{ij}=A_{ji}$ (unoriented links). The matrix stochastic process
$A(t)$ in continuous time $t$ represents the evolution of the network.

Two local parameters will be of central importance for our discussion namely
the node degree $c_{i}(t)=\sum_{j}A_{ij}(t)$ and the local clustering
coefficient
\begin{equation}
q_{i}(t)=\frac{\sum_{j<k}A_{ik}A_{ij}A_{jk}}{\sum_{j<k}A_{ik}A_{ij}}.
\end{equation}%
The latter measures the fraction of pairs of neighbors of $i$ who are also
neighbors among themselves. The averages of these two quantities over sites
will be simply denoted $c$ and $q$. While random networks have $q\sim 1/N$,
social networks typically have a clustering coefficient \cite{wa_stro_98}
bounded above zero.

Denote by $F_{i}=\{j|A_{ij}=1\}$ the set of neighbors (\textquotedblleft
friends\textquotedblright ) of the node $i$. The network evolves due to the
following three processes.

1. \emph{Long distance search}: At rate $\eta $, each node $i$ gets the
opportunity to make a link to a node $j$ randomly selected (if the link is
already there nothing happens).

2.\emph{\ Short distance search}: At rate $\xi ,$ each node $i$ picks
at random one of its neighbors $j\in F_{i}$ and $j$ then randomly
selects (i.e.  \textquotedblleft refers to\textquotedblright ) one of
its other neighbors $k\in F_{j}\backslash \{i\}$. If $k\not\in F_{i}$
then the link between $i$ and $k$ is formed. If $F_{i}=\emptyset $ or
$F_{j}=\{i\}$ or $k\in F_{i}$ nothing happens.

3. \emph{Decay}: At rate $\lambda ,$ each existing link decays and it
   is randomly deleted.

\begin{figure}[tbph]
%\vspace*{60mm} %
\includegraphics[width=80mm]{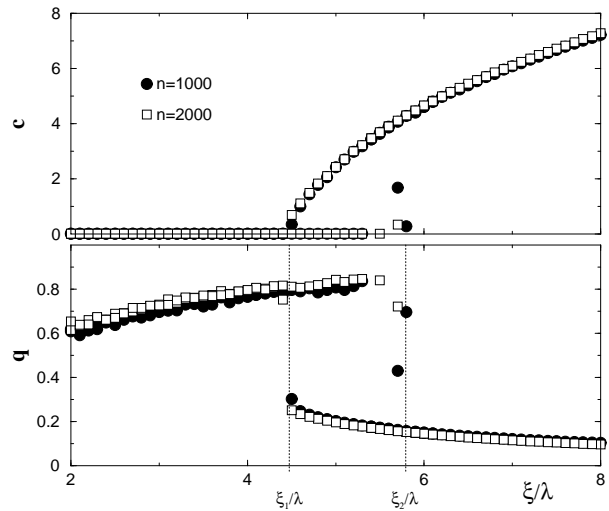}
\caption{Average degree $c$ (top) and clustering coefficient $q$ (bottom)
from numerical simulations with $\protect\eta /\protect\lambda =0.01$ for
populations of size $n=1000$ and $2000$. Runs were equilibrated for a time 
$t_{\mathrm{eq}}=3000/\protect\lambda $ before taking averages.}
\label{figsim}
\end{figure}

For $\xi =0,$ the dynamics is very simple and the stationary network is a
random graph with average degree $c=\eta /\lambda $. For $\eta \ll \lambda $
the network is composed of many disconnected parts. Fig. \ref{figsim}
reports what happens when the local search rate $\xi $ is turned on. For
small $\xi ,$ network growth is limited by the global search process that
proceeds at rate $\eta $. Clusters of more than $2$ nodes are rare and when
they form local search quickly saturates the possibilities of forming new
links. Suddenly, at a critical value $\xi _{2}$, a giant component
connecting a finite fraction of the nodes emerges. The average degree $c$
indeed jumps abruptly at $\xi _{2}$. The distribution $p(c)$ of $c_{i}$ is
peaked with an exponential decrease for large $c$ and a power law $p(c)\sim
c^{\mu }$ for $c$ small. The network becomes more and more densely connected
as $\xi $ increases further. But when $\xi $ decreases, we observe that the
giant component remains stable also beyond the transition point ($\xi <\xi
_{2}$). Only at a second point $\xi _{1}$ does the network lose stability
and the population gets back to an unconnected state. There is a whole
interval $[\xi _{1},\xi _{2}]$ where both a dense-network phase and one with
a nearly empty network coexist. This behavior is typical of first-order
phase transitions. The coexistence region $[\xi _{1},\xi _{2}]$ shrinks as 
$\eta$ increases and it disappears for $\eta >0.05\lambda $.

The average clustering coefficient $q$ shows a non-trivial behavior. In the
unconnected phase, $q$ increases with $\xi $ as expected. In this phase, $q$
is close to one because the expansion of the network is mostly carried out
through global search, and local search quickly saturates all possibilities
of new connections. On the other hand, in the dense-network phase, $q$ takes
relatively small values. This makes local search very effective. Remarkably
we find that $q$ \emph{decreases} with $\xi $ in this phase, which is rather
counterintuitive: \emph{increasing the rate }$\xi $\emph{\ at which bonds
between neighbors form through local search, the density $q$ of these bonds
decreases}. In fact, similar behavior is found if, fixing $\xi $ and $\eta ,$
the volatility rate $\lambda $ decreases.

To shed light on these numerical results, we study the dynamics of the
distribution $P(\vec{c},\vec{q},t)$ of the degrees $\vec{c}$ and clustering
coefficients $\vec{q}$. Specifically, we study a mean field approximation
that assumes $q_{i}=q$ for all $i=1,\ldots ,N$ and

\begin{equation}
P(\vec{c},\vec{q},t)=\prod_{i=1}^{N}p(c_{i},t)\delta (q_{i}-q).
\label{mfassump}
\end{equation}

\noindent It is convenient to set $\lambda =1,$ by an appropriate time
rescaling. Then, the transition rates that enter into the master equation
for $p$ have the form:

\begin{eqnarray}
w(c\rightarrow c+1) &=&2\eta +\beta \theta(c) +\gamma c  \label{growth} \\
w(c\rightarrow c-1) &=& c  \label{decay}
\end{eqnarray}

\noindent where $\theta (k)=0$ for $k\leq 0$ and $\theta (k)=1$ for $k>0$.
In Eq. (\ref{growth}), the term $2\eta$ accounts for long distance search.
The factor $2$ counts both the processes when the search opportunity is
given to site $i$ and when it is given to another site, which selects $i$ as
end point. The second term in Eq. (\ref{growth}) arises from local search
and it requires that $c_i>1$. Here $\beta=\xi(1-q)P\{c_j>1|j\in F_i\}$
accounts for the fact that the selected friend $j\in F_i$ must have at least
one more friend $k\neq i$ and that $k\not\in F_i$, which occurs with
probability $1-q$.

Finally the last term accounts for indirect local search opportunities given
to a friend $k$ of a friend $j\in F_{i}$ of $i$. This process is
proportional to $c_{i}$ and $\gamma =\xi (1-q)\langle {c_{k}^{-1}}\rangle $
accounts for the probability that $i\not\in F_{k}$ and $k$ selects $j$. 
Note that the probability that $j\in F_k$ selects $i$ is
$1/(c_j-1)$. This, combined with the multiplicity $c_j-1$ of second
neighbors $k\in F_j$ of $i$, contributes a factor
$(c_{j}-1)/(c_{j}-1)=1$ to the rate.

Both $\beta$ and $\gamma$ will be determined self-consistently. The master
equation for $p(c)$, in the stationary state, can be solved using the
generating function $\pi(s)=\langle{s^c}\rangle$:

\begin{equation}
\pi (s)=\frac{\beta +2\eta (1-\gamma s)^{-\mu }}{\beta +2\eta (1-\gamma
)^{-\mu }},~~~~\mu =\frac{2\eta +\beta }{\gamma }.  \label{pis}
\end{equation}%
Note that $p(c)\sim c^{\mu }$ for small $c$ and $p(c)\sim e^{-|\ln \gamma
|c} $ for large $c$ which perfectly matches the behavior observed in
numerical simulations.

Eq. (\ref{pis}), allows us to compute the distribution
$P\{c_{j}=k|j\in F_{i}\}=\tilde{p}(k)$ for the degree $c_{j}$ of $j\in
F_{i}$. The larger $c_{j}$ the more likely is $j$ a neighbor of
$i$. Thus, $\tilde{p}(k)\propto kp(k)$ which, in terms of the
generating function, implies $\tilde{\pi}(s)=s\pi ^{\prime }(s)/\pi
^{\prime }(1)$. Using this to compute averages over $c_{j}$ and
$c_{k}$ we arrive at the self-consistent equations:

\begin{eqnarray}
\beta &=&\xi (1-q)\left[ 1-\frac{\pi ^{\prime }(0)}{\pi ^{\prime }(1)}\right]
\label{selfconsb} \\
\gamma &=&\xi (1-q)\frac{1-\pi (0)}{\pi ^{\prime }(1)}  \label{selfconsg}
\end{eqnarray}

\noindent As we should, in the limit $\xi \to 0$ we find $\beta,\gamma\to 0$
and we recover a pure Poisson distribution with mean $2\eta$. But with
constant $q$ Eqs. (\ref{selfconsb},\ref{selfconsg}) are not able to
reproduce the observed behavior. It just predicts a smooth crossover and no
phase coexistence. This means that, in order to shed light on our
observations, it is essential to allow for $q$ to depend on the parameters
of the model and the distribution $p(k)$.

In order to derive an equation for $q$, we consider the evolution of
clustering for a node $i$ with $c_{i}$ friends. Let $Q_{i}=
\sum_{j<k}A_{ij}A_{ik}A_{jk}=q_{i}c_{i}(c_{i}-1)/2$ be the number of pairs
of friends of $i$ which are also friends among themselves. Only local search
processes contributes to an increase in $Q_{i}$ through two different routes.

The first is when a local search opportunity is given to site $i$
itself and has already been discussed above. Its rate is
$W_{1}(Q_{i}\rightarrow Q_{i}+1)=\xi (1-q)P\{c_{i}>1\}$. The second
occurs when a local search opportunity is given to some $j\in F_{i}$,
who then asks to $i$ about his other friends $k\in F_{i}$ ($k\neq
j)$. This may lead to the formation of the link between $j$ and $k$,
thus increasing $Q_{i}$ by one.  The rate of this process is given by
$W_{2}(Q_{i}\rightarrow Q_{i}+1)=\xi \langle {c_{i}\theta
(c_{i}-1)}\rangle \langle {c_{j}^{-1}}\rangle (1-q)$.  Here, $1/c_{j}$
is the probability that $j$ picks $i$ from his neighbors and $1-q$ is
the probability that $k\not\in F_{j}$. This rate should is multiplied
by the number $c_{i}$ of neighbors of $i$, but is zero unless
$c_{i}\geq 2$~\cite{note}. Finally, we must account for the link-decay
process that, contrary to the former two, decreases $Q_{i}$. The rate
at which this happens is $W_{\lambda }(Q_{i}\rightarrow
Q_{i}-1)=Q_{i}=\left\langle c_{i}(c_{i}-1)\right\rangle q/2$.

If we now take the averages on $c_i$ and $c_j$ with probability
distributions $p(k)$ and $\tilde p(k)$ respectively, we can impose
stationarity on $Q_{i}$, i.e. $\langle {\Delta Q_{i}}\rangle
=W_{1}+W_{2}-W_{\lambda }=0$. After some algebra, this condition becomes the
desired equation for $q$:
\begin{equation}
\frac{q}{2}\pi^{\prime\prime}(1)=\xi(1-q)[2-\pi(0)]\left[1-
\frac{\pi^{\prime}(0)}{\pi^{\prime}(1)}\right]  \label{q}
\end{equation}

Using Eq. (\ref{pis}), we arrive at a set of three self-consistent equations
for $\beta ,~\gamma $ and $q$. %
%%%start figure 3
\begin{figure}[tbph]
%\vspace*{60mm} %
\includegraphics[width=80mm]{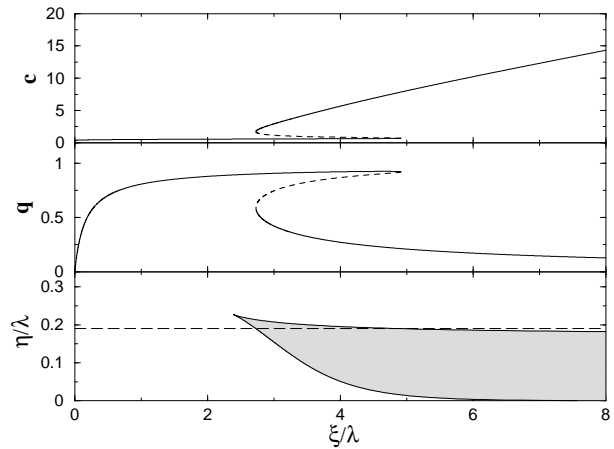}
\caption{Mean degree $c$ (top) and clustering coefficient $q$ (middle) as a
function of $\protect\xi /\protect\lambda $ for $\protect\eta /\protect%
\lambda =0.19$. Bottom: phase diagram in the mean field approximation. The
behaviour of $c$ and $q$ along the dashed line is reported above.}
\label{mfres}
\end{figure}
%%%end figure 3

The solution of Eqs. (\ref{selfconsb}, \ref{selfconsg}) and (\ref{q})
is shown in Fig. \ref{mfres} for $\eta =0.19$. The average degree
$c=\pi ^{\prime }(1)$ and clustering coefficient $q$ exhibit the same
dependence on $\xi $ as that found in numerical simulations. Depending
on the values of $\xi $ and $\eta $ we find either a unique solution
with small $c$ and large $q$ (corresponding to a dilute network) or a
unique solution with large $c$ and small $q$ (dense network), or both
solutions simultaneously. In particular the solution correctly
reproduces the behavior of $q$ in the two phases: $q$ increases with
$\xi $ in the dilute network phase whereas it decreases with $\xi $
when a giant component forms. Our approach shows that this is not just
a by-product of our analysis but rather an essential ingredient for
understanding the network's dynamics. Fig. \ref{mfres} also depicts
the phase diagram predicted by the mean field. In the shaded region,
Eqs. (\ref{selfconsb}, \ref{selfconsg}, \ref{q}) have three solutions,
of which two are stable and correspond to the two coexisting phases.

The coexistence interval $[\xi_1,\xi_2]$ predicted by the mean field
theory is much larger than that observed in numerical simulations, as
can be seen by comparing Fig. \ref{figsim} and Fig. \ref{mfres}. For
example, the critical point $\eta_c/\lambda\cong 0.226\ldots$ above
which there is a smooth crossover in the mean field is an order of
magnitude larger than that suggested by numerical simulations. We
believe this is due to the fact that mean field theory
underestimates fluctuations and neglects correlations.

It is straightforward to repeat our approach to obtain the mean field
equations for the model of Ref. \cite{DEB}. The network growth process
of Ref. \cite{DEB} mixes local and global search in a different
way. (For example, search is always effective since when it is
unsuccessful locally, the agent nevertheless creates a link through
global search.) In addition volatility affects sites instead of
removing bonds, i.e. Eq. (\ref{decay}) is replaced with
$w(c_{i}\rightarrow 1)=p$. This changes considerably the stationary
state distribution, since for $c>1$ the stationarity in the master
equation implies $p(c)=\frac{a+c-1}{a+p/\gamma +c}p(c-1)$, i.e. a
power law behavior $p(c)\sim c^{-p/\gamma -1}$, as observed in
Ref. \cite{DEB}. The solution of the self-consistent equations is
always unique, implying that there no phase transition in this
model. These conclusions illustrate that the present approach is a
rather powerful tool in the analysis of network dynamics in a wide
range of different setups.

To sum up, our aim in this paper has been to study a simple model of network
formation whose implications shed some light on the evolution (rise and
fall) of a networked society. The induced network dynamics displays a
first-order transition that, as the environmental conditions improve, lead
from a sparse phase to a qualitatively different regime where the rich
potential of a network society is realized. Thus, in particular, social
interaction becomes dense (i.e. average connectivity is high) and individual
search turns effective (i.e. redundant search is avoided by low clustering).

These findings explains the apparently paradoxical observation that a
networked society does not necessarily materialize even under favorable
conditions while, by contrast, it displays a significant resilience to
deteriorating conditions. This may help understand the origin of the
\textquotedblleft miracles\textquotedblright\ and \textquotedblleft
anti-miracles\textquotedblright\ in economic development \cite{L}, which are
still an unresolved puzzle for modern economic theory.

%\begin{acknowledgments}
This work was supported by EXYSTENCE network of excellence of EU, the
Spanish Ministry of Education (BEC2001-0980), the Grant Agency of the
Czech Republic, project No. 202/01/1091 and by the EU grant
HPRN-CT-2002-00319.
%\end{acknowledgments}


\begin{thebibliography}{99}
\bibitem{note} There is a third process that may give $\Delta
Q_{i}=+1$.  This occurs when $j\in F_{i}$ is given a local search
opportunity, then a site $\ell \neq i$ is picked as intermediate node
and $k\in F_{\ell }\cap F_{i}$ is selected to form a link to $j$. It
is hard to determine the joint probability of these events precisely
because fourth-order correlations (i.e. four sites) are involved. In
the spirit of the mean field approach, we may neglect it. In any case,
it is easy to check that $W_{2}(Q_{i}\rightarrow Q_{i}+1)$ is a safe
upper bound to the rate of this process. We could therefore account
for this fact by introducing a factor $\alpha \in \lbrack 1,2]$ in
front of $W_{2}$ above. However, since the final results depend very
weakly on $\alpha $, we choose to set $\alpha =1$ and ignore this last
process altogether.
\end{thebibliography}
\end{document}